\documentclass[preprint]{aastex63}
\pdfoutput=1
\received{2020 March 29}
\revised{2020 June 4}
\accepted{2020 June 5}
\submitjournal{ApJS}

\shorttitle{The SOHO/EIT Flare Catalog}
\shortauthors{rotti, martens and aydin}
\graphicspath{{./}{figures/}}
\begin{document}

\title{A Catalog of Solar Flare Events Observed by the SOHO/EIT}

\correspondingauthor{Sumanth A. Rotti}
\email{rotti@astro.gsu.edu}

\author[0000-0003-1080-3424]{Sumanth A. Rotti}
\affiliation{Department of Physics and Astronomy \\
Georgia State University \\
Atlanta, GA 30303, USA}

\author{Petrus C.H. Martens}
\affiliation{Department of Physics and Astronomy \\ 
Georgia State University \\
Atlanta, GA 30303, USA}

\author[0000-0002-9799-9265]{Berkay Aydin}
\affiliation{Department of Computer Science \\ 
Georgia State University \\
Atlanta, GA 30303, USA}



\begin{abstract}

We have compiled a catalog of solar flares as observed by the Extreme ultraviolet Imaging Telescope (EIT) aboard the Solar and Heliospheric Observatory (SOHO) spacecraft and the GOES spacecraft over a span from 1997 to 2010. During mid-1998, the cadence of EIT images was revised from \textit{two images per day} to \textit{12 minutes}. However, the low temporal resolution causes significant data gaps in capturing much of the flaring phenomenon. Therefore, we monitor possible errors in flare detection by flare parameters such as temporal overlap, observational wavelength, and considering full field of view (FOV) images. We consider the GOES flare catalog as the primary source. We describe the technique used to enhance the GOES detected flares using the Extreme Ultraviolet (EUV) image captured by the EIT instrument. In order to detect brightenings, we subtract the images with a maximum cadence of 25 minutes. We have downloaded and analyzed the EIT data via the Virtual Solar Observatory (VSO). This flare dataset from the SOHO/EIT period proves indispensable to the process of the solar flare predictions as the instrument has covered most of Solar Cycle 23.

\end{abstract}



\section{Introduction} \label{sec:intro}

Solar flares (SFs) are sudden brightening features across the electromagnetic spectrum observed due to the release of stored magnetic energy in the solar atmosphere \citep{Benz2016}. Such an instant explosion of energy is followed by a release of intense radiation into space. SFs are one of the most prominent activities that are often accompanied by coronal mass ejections (CMEs). However, not all flares have accompanying CMEs. Larger CMEs lead to solar energetic particle (SEP) emissions. These high-energy particles consist of electrons, protons, and higher ions accelerated to near-relativistic speeds. Often, SEP events are classified as gradual and impulsive where the latter focus on SFs while the former are relevant to those SEPs originating from CMEs \citep{koul,desai,kahler}. Even though SEPs are associated with large flares, not every large flare has SEPs associated with it. In this regard, the X-ray peak contends as a strong indicator for a large SF that is capable of accelerating high-energy particles toward the Earth.

In space weather (SWx) research, the solar events are studied for (i) their influences on the nature of geomagnetic fields and (ii) how they can interact with the electromagnetic systems of the Earth \citep{pevtsov2016}. Some extreme events affect the radiation environment traveled by many spacecraft and astronauts outside the Earth's magnetosphere, posing an extreme threat to humans on space missions \citep{dartnell_2011, melott}. High-energy rays from SFs can cause disturbances in the upper atmosphere of the Earth at $\approx 80$ km, which can cause radio blackouts. The degradation of high-frequency radio signals occurs due to ionization in the D region of a more dense and lower layers of the ionosphere, primarily impacting the 3-30 MHz band \citep{benson, namara, frissell} This includes affecting certain types of global military communications and air traffic control \citep{doi:10.1029/2019SW002254}.

SFs are monitored continuously in white light, radio waves, soft and hard X-rays, and in extreme ultraviolet (EUV) wavelengths by several instruments on the ground and in space \citep{Benz2016}. The \textit{Geostationary Operational Environmental Satellites}\footnote{\url{https://www.ngdc.noaa.gov/stp/solar/solarflares.html}} (GOES, \citet{goes}) have collected a database of solar flares providing temporal and spatial data along with intensity classifications. Most of the solar missions such as, YOHKOH\footnote{\url{http://ylstone.physics.montana.edu/ylegacy/HXT_catalog/index.html}} \citep{yohkoh}, the \textit{Reuven Ramaty High Energy Solar Spectroscopic Imager}\footnote{\url{https://hesperia.gsfc.nasa.gov/hessidata/dbase/hessi_flare_list.txt}} (RHESSI, \citet{rhessi}), HINODE\footnote{\url{http://xrt.cfa.harvard.edu/flare_catalog/}} \citep{Kosugi2007}, and the \textit{Solar Dynamics Observatory} (SDO)\footnote{\url{https://www.lmsal.com/isolsearch}} (\citep{AIA}) have their flare catalogs, whose events are curated using onboard instruments. These catalogs use GOES X-rays mostly to detect flares but imaging from other ones to localize the flare.

The metadata obtained from various observational instruments has been processed and produced in the form of a catalog by many researchers for scientific study essential to solar physicists and space weather researchers (e.g., Yohkoh Flare Catalog: \citet{yflare}, Hinode Flare Catalog: \citet{hinodeflare}, SDO/EUV Variability Experiment (EVE) Catalog: \citet{hock}).  There are also efforts by \citet{Sadykov_2017} in combining metadata from various instruments into one database useful for predictive analytics. The integration of data as well supports researchers alongside the technology of Big Data.\footnote{Interactive Multi-Instrument Database of Solar Flares \url{https://helioportal.nas.nasa.gov/}} To analyze flares, the data from a specific instrument and observation information are essential for statistical as well as analytical studies. For example, an active region (AR) producing a GOES $>$M1.0 flare has a larger probability of producing flares in the next 24 hr \citep{park}. The location of the parent SF is an essential parameter for the efficient prediction of a possible SEP event \citep{gopal}.

Since 1974, two X-ray sensors (XRS) on each GOES satellites measure solar X-ray fluxes for the wavelength bands of 0.5-4 {\AA} (short channel) and 1-8 {\AA} (long channel; \citep{garcia94}. The flares from GOES are detected through hard X-ray detectors and provides the flare location on the Sun using the H-alpha images and the Solar X-ray Imager instrument uninterrupted from 2003 \citep{goes}. GOES has traced flaring activity since the beginning of 1974 and has a catalog available with spatial and temporal specifications along with an associated AR number. However, some flare locations and/or AR numbers are missing from the GOES catalog or probably have errors \citep{milligan,berkay,dml}.

Amongst several operational instruments, the \textit{Extreme ultraviolet Imaging Telescope} (EIT) on board the \textit{Solar and Heliospheric Observatory} (SOHO) observed the Sun in EUV wavelengths \citep{EIT}. Although EIT was not designed to observe flares, SOHO did capture a large number of flares during Solar Cycle 23. Nonetheless, no catalog of flares has been available based on the EIT data until now. Although our work is not exclusive to EIT, we are enhancing the GOES detected flares using EIT data. The enhancement is rather the important localization aspect as it increase the quality of the flare database. Such a catalog would be very useful to researchers to investigate more for statistical and other SF studies further.
 
 In this paper, we present a catalog of SFs, as seen by the SOHO/EIT instrument. This part of our work is introduced in Section \ref{sec:data}. We choose the duration of the EIT's primary operations, i.e., from 1997 to 2010, to get the flare locations. The main purpose behind this effort is to improve the quality of the flare locations in the existing database curated for machine learning (ML) and to make the flare metadata during the SOHO-era  available for interested researchers. We have examined all the spatiotemporal attributes and/or characteristics of the EIT flares alongside the GOES SF catalog that is explained with the results in Section \ref{sec:method}. We summarize our results in Section \ref{sec:conc}. Our catalog will be made publicly available on Harvard dataverse at \url{https://doi.org/10.7910/DVN/C9H34R}.

\section{DATA} \label{sec:data}

In addition to the traditional statistical survey, for ML-based SWx forecasts, it is crucial to verify and establish the input data in the GOES catalog with other space-borne instruments \citep{berkay}. Flare prediction is one of the most prominent SWx forecasting applications for which flare localizations are of utmost importance. In this regard, \citet{dml} developed a benchmark dataset of ARs, together with flares consisting of a clean, verified, and easily accessible database put together from several sources using SDO data. In order to extend the quality of this benchmark dataset, we choose to analyze the pre-SDO period utilizing the SOHO data.

\subsection{EIT Imaging} \label{subsec:imaging}

Level Zero science data from the EIT instrument was downloaded by querying the \textit{Virtual Solar Observatory} (VSO)\footnote{\url{https://vso.nascom.nasa.gov/cgi/search}} using the Solarsoft\footnote{\url{http://www.lmsal.com/solarsoft/ssw_whatitis.html}} IDL routines. The EIT images are already corrected for pointings in the header. The EIT captures the Sun in four different EUV wavelengths, which are, 171, 195, 284, and 304 \AA. The EIT instrument has a Nyquist limited resolution of 5.2 arcsec (2.6 arcsec pixels) \citep{EIT}. The initial cadence for observing was six hours but was improved to 12 minutes from July 1998 on \citep{esa}. Nonetheless, the available 195 {\AA} synoptic images have a varying cadence from seven minutes to 13 minutes.

\begin{figure}%[ht!]
    \centering
    \includegraphics[width=1\textwidth,trim={5cm 5cm 5cm 5cm},clip]{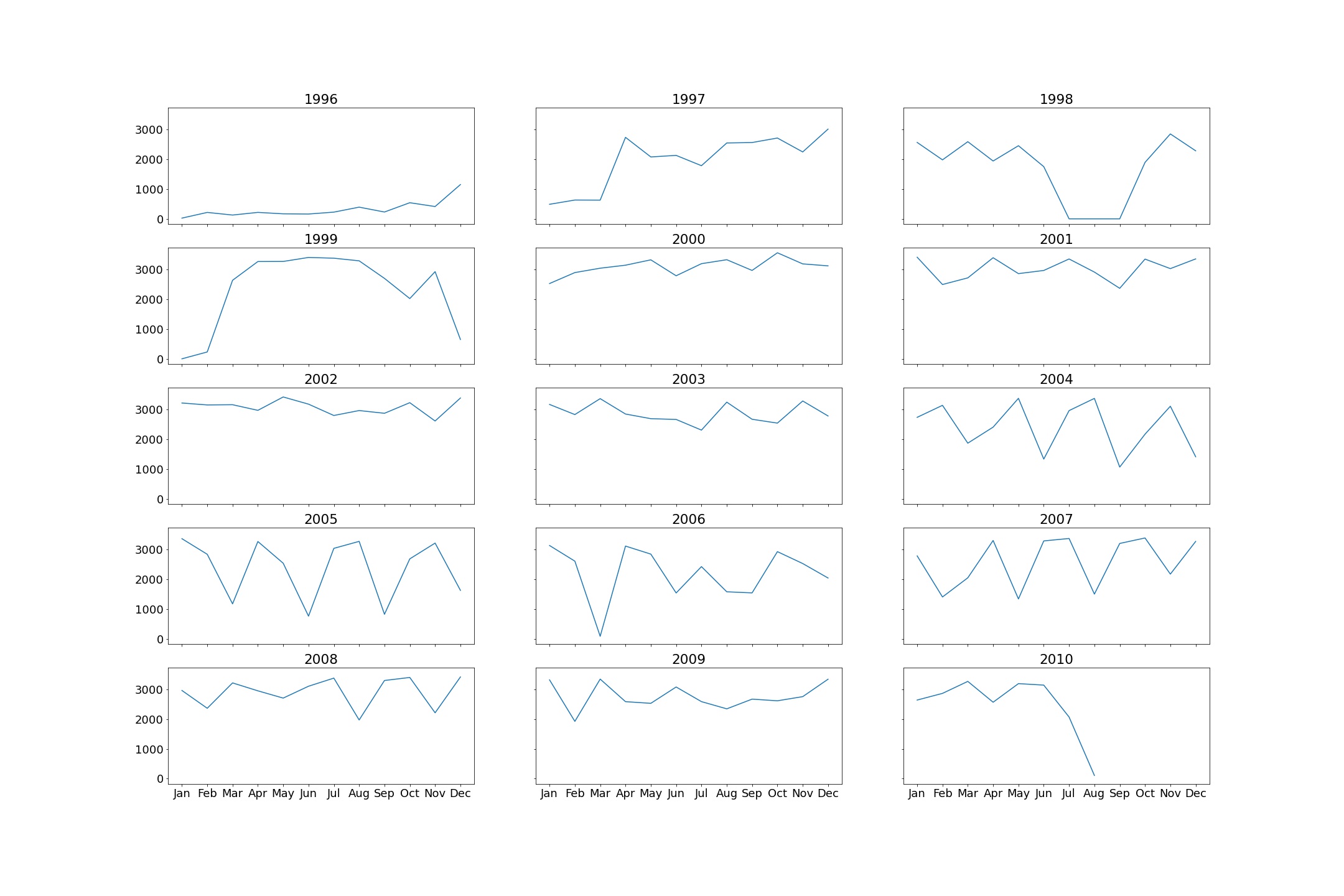}
    \caption{Number of SOHO/EIT images available per month from 1996 to 2010.}
    \label{fig:data}
\end{figure}

In Figure \ref{fig:data}, the number of available EIT images per month from 1996 to 2010 is shown. At the specified cadence of 12 minutes, there should be approximately 3600 images captured by EIT every month. However, that is not the case. Noticeably, there is a significant irregularity in the available EIT images, which is a primary challenge for finding flare locations. There is an intrinsic difference between X-ray observations and those in the EUV images and they do not necessarily correspond one-to-one with each other. In the 195 {\AA} band, Fe XII, Ca XVII, and Fe XXIV lines are dominated, which have their maximum fractional abundances at 1.4 $\times$ $10^6$ K, 5 $\times$ $10^6$ K, and (1–2)$\times$ $10^7$ K, respectively \citep{Feldman_1999}. The 195 {\AA} wavelength shows significantly hot flaring regions on the Sun, giving a better choice to detect flares. However, 304 {\AA} also shows the transition region of the solar atmosphere, effectively reducing the contrast between flares and the background Sun compared to the other three wavelengths, and therefore was not considered in this work.

\subsection{Data selection} \label{subsec:dselect}

To localize the flares and associate them with ARs, three data resources are needed: (1) the GOES flare catalog\footnote{\url{https://www.ngdc.noaa.gov/stp/space-weather/solar-data/solar-features/solar-flares/x-rays}}, (2) the \textit{National Oceanic and Atmospheric Administration} (NOAA) AR summary, and (3) the full field-of-view EIT images. For our study, we used 195 {\AA} images of the EIT instrument with a resolution of 1024 x 1024 pixels. The \textit{EIT Web catalog interface} and VSO hold the data products from SOHO/EIT, which were downloaded to our server. More details about the instrument and calibration can be obtained from the official website.\footnote{\url{https://umbra.nascom.nasa.gov/eit/eit_guide/}}

\section{Methodology} \label{sec:method}

Flare detection schemes are often built upon (1) time series data and/or (2) image data, for automatic detection \citep{borda, qu, kano, grigis, aranda, bonte, cab, refId0}. For this work, we utilized Solarsoft's\dataset[\textit{Flare Locator}]{https://hesperia.gsfc.nasa.gov/ssw/gen/idl/ssw_util/ssw_flare_locator.pro} algorithm \citep{freeland} that works on the second principle-that is, image-based data. The flares are detected by measuring a threshold of the peak light intensity per macro pixel of an EIT 195 {\AA} image closest to the GOES start time and then taking the relative difference with an image closest to the GOES peak time. \citet{milligan} report that this process of image subtraction does
not actually guarantee a flaring emission is detected but only that the timing and pointing given in GOES dataset are consistent with the timing and location of the flare. The image subtraction using \textit{Flare Locator} algorithm also extracted information about dynamic flaring events in the EIT data that are not seen in X-ray images and, hence, cannot be related to a GOES flare. These are not false reports but are multiple events or, in some cases, post-flare loops occurring over a different active region or part of the Sun. Timing and position information from each event can be cross-checked with the metadata from other instruments to determine the observation of these events at the same time in the same location. The procedure to obtain the input images to the \textit{Flare Locator} is discussed below.

\subsection{Flare detection and verification} \label{subsec:fdv}

The steps followed in preparing the EIT Flare Catalog include:
\begin{enumerate}
    \item All the flares reported in the GOES catalog are downloaded from the NOAA site during SOHO/EIT's primary operation from 1997 to 2010.
    \item For each flaring event of GOES class $>$ C1.0, a full-disk  EIT image at 195 {\AA} is identified, such that the imaging time corresponds to the GOES \textit{start time} of the flare within the range of $\pm$ 5 minutes.
    \item Every available image is then input as a primary image to the \textit{Flare Locator} with a reference image taken after the primary within the range $\pm$ 25 minutes. If multiple images are available within the considered time range, then the closest image observed after the GOES flare \textit{peak time} is selected as the reference image.
\end{enumerate}

\vspace{5mm}

The verification process between the EIT flare coordinates and GOES flare coordinates is two-fold based on the following convention:
\begin{enumerate}
\item Exclusive flare positions in the GOES catalog are compared with the EIT flare coordinates such that the flares overlap in their temporal occurrence.
\item The implicit location derived from the centroid of NOAA AR is used if the explicit point coordinates are not available
\end{enumerate}

\vspace{5mm}
In Table \ref{tab1}, a sample dataset of verified X-class flares from 1997 to 2000 is shown. The dataset provides the index, start and end times of the event detection, location of the flaring region, the GOES X-ray classification, and the NOAA AR number. The following convention is used to determine whether a flare is flagged as \textit{verified} or \textit{not verified}:
If the flare position (X, Y) in the EIT image is within 25 arcsec compared to the GOES location, then it is considered \textit{verified}. The selection of the distance is based on a similar convention implemented by the Hinode flare catalog to identify the overlapping flaring regions \citep{hinodeflare}. If the EIT flare coordinates within $\pm$ 25 arcsec do not overlap with the GOES flare catalog, then they are considered as \textit{not verified}. There are some events detected in the EIT images where the corresponding locations vary largely from GOES coordinates. This is due to co-occurring events around the same time on different parts of the Sun. Such events are also flagged as \textit{not verified}. The \textit{not verified} class would include the flares occurring near-limb or behind the limb that would be detected by GOES X-rays as a result and if there is no way here to localize those events. If there are no existing coordinates in the GOES catalog to cross-check the EIT coordinates, then such flares are also flagged as \textit{not verified}. The structure of the database schema-that is, the headers in Table \ref{tab1}-are as follows:

\begin{itemize}
\item start\_time\_detection corresponds to the observation time of the initial EIT image used for detection. This is the primary image input to the flare detection software.
\item end\_time\_detection corresponds to the observation time of the preceding EIT image used for detection where the second image is input as a reference to obtain the difference image.
\item eit\_fl\_location is the value of flare locations in Heliographic Stonyhurst (HGS) coordinates where {\tt\string north} and {\tt\string west} is positive while {\tt\string south} and {\tt\string east} is negative. The number of decimal places here is limited to two values.
\item goes\_class represents the classification of the flare magnitude according to GOES standards.
\item noaa\_active\_region is a number assigned to the flaring region by the NOAA group available in the GOES catalog. Here, N/A refers to not available AR numbers in the GOES catalog.
\end{itemize}

The dataset consists of a 'Remarks' column specifying whether the flare is verified or not. If not verified, then the co-occurring events or those with unavailable GOES location are specified correspondingly to maintain distinction. For such events, the GOES classification column in the EIT catalog retains the original values of that flare from the GOES catalog.

I\begin{deluxetable*}{cccccc}
\tablenum{1}
\tablecaption{EIT flare list of GOES X-classes from 1997 to 2000. \label{tab1}}
\tablewidth{0pt}
\tablehead{
\colhead{Index} & \colhead{start\_time\_detection} & \colhead{end\_time\_detection} & \colhead{eit\_fl\_location} &
\colhead{goes\_class} &  \colhead{noaa\_active\_region}  \\ 
\colhead{} & \colhead{} & \colhead{} & \multicolumn1c{(hgc degrees)} & \colhead{} & \colhead{}
}
\startdata
1 & 1997-11-04T05:41 & 1997-11-04T05:58 & (30.50,-17.96) & X2.1 & 8100 \\
2 & 1997-11-6T11:41 & 1997-11-06T12:01 & (61.00,-19.57) & X9.4 & 8100 \\
3 & 1997-11-27T13:12 & 1997-11-27T13:38 & (-62.77,15.29) & X2.6 & 8113 \\
4 & 1998-04-23T05:35 & 1998-04-23T05:51 & (70.66,19.47) & X1.2 & N/A \\
5 & 1998-04-27T09:21 & 1998-04-27T09:36 & (-53.24,-16.70) & X1.0 & 8210 \\
6 & 1998-05-02T13:21 & 1998-05-02T13:42 & (9.55,-15.04) & X1.1 & 8210  \\
7 & 1998-05-06T08:10 & 1998-05-06T08:23 & (65.46,-15.81) & X2.7 & N/A \\
8 & 1998-11-22T06:29 & 1998-11-22T06:44 & (55.16,-27.63) & X3.7 & N/A \\
9 & 1998-11-22T16:09 & 1998-11-22T16:23 & (65.30,-27.97) & X2.5 & N/A \\
10 & 1998-11-23T06:29 & 1998-11-23T06:44 & (69.10,-28.12) & X2.2 & 8384  \\
11 & 1998-11-24T02:12 & 1998-11-24T02:24 & (78.82,-29.50) & X1.0 & N/A \\
12 & 1998-11-28T05:12 & 1998-11-28T05:48 & (-53.08,20.03) & X3.3 & N/A \\
13 & 1999-08-02T21:24 & 1999-08-02T21:36 & (46.53,-17.45) & X1.4 & 8647 \\
14 & 1999-08-28T18:00 & 1999-08-28T18:12 & (13.93,-29.55) & X1.1 & 8674 \\
15 & 1999-11-27T12:00 & 1999-11-27T12:12 & (62.70,-14.99) & X1.4 & 8771 \\
\enddata
\end{deluxetable*}

\begin{figure}[ht!]
    \centering
    \includegraphics[width=1.0\textwidth]{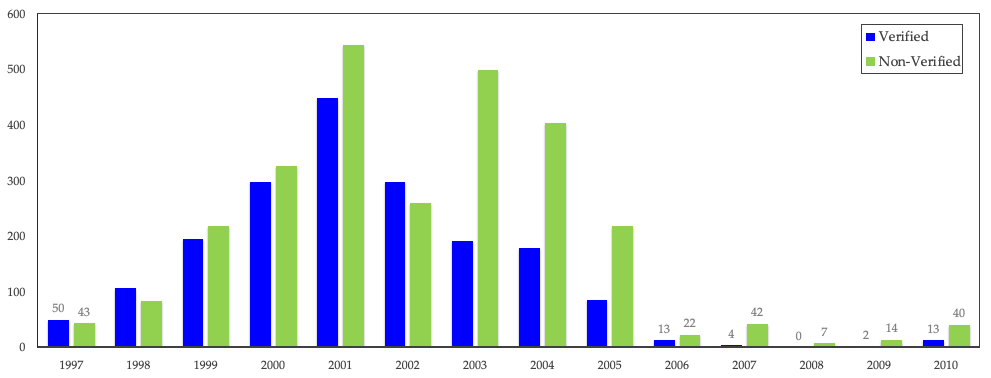}
    \caption{A bar plot of all the verified and non-verified flares of the SOHO/EIT from 1997 to 2010.}
    \label{fig:4}
\end{figure}

\begin{figure}[ht!]
    \centering
    \includegraphics[width=1.0\textwidth]{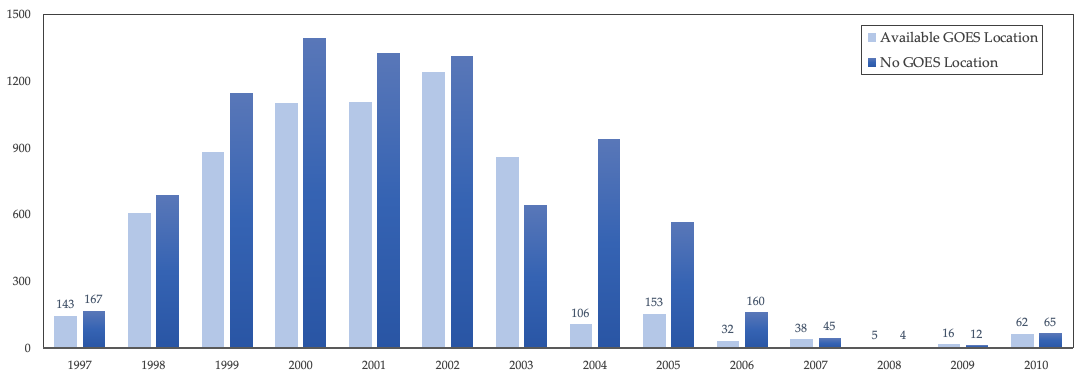}
    \caption{A bar plot showing the number of available and missing flare locations in the GOES catalog from 1997 to 2010.}
    \label{fig:9}
\end{figure}
\begin{figure}[ht!]
  \centering
    \includegraphics[width=1.0\textwidth]{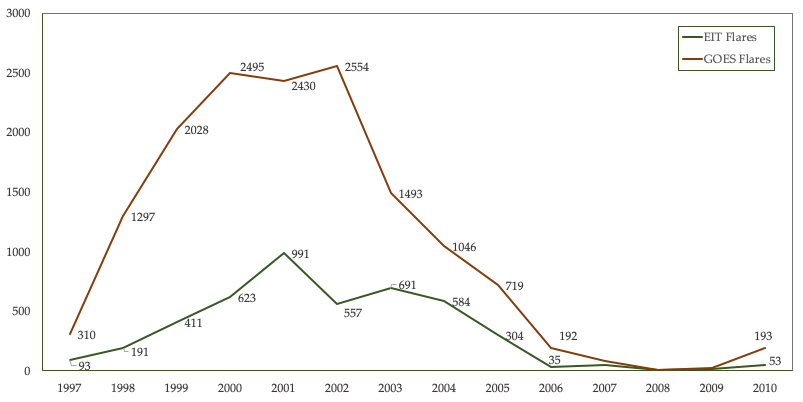}
    \caption{The EIT and GOES flares from 1997 to 2010.}
    \label{fig:7}
\end{figure}

\begin{figure}[ht!]
    \centering
    \includegraphics[width=0.9\textwidth,trim={0 8cm 0 8.5cm},clip]{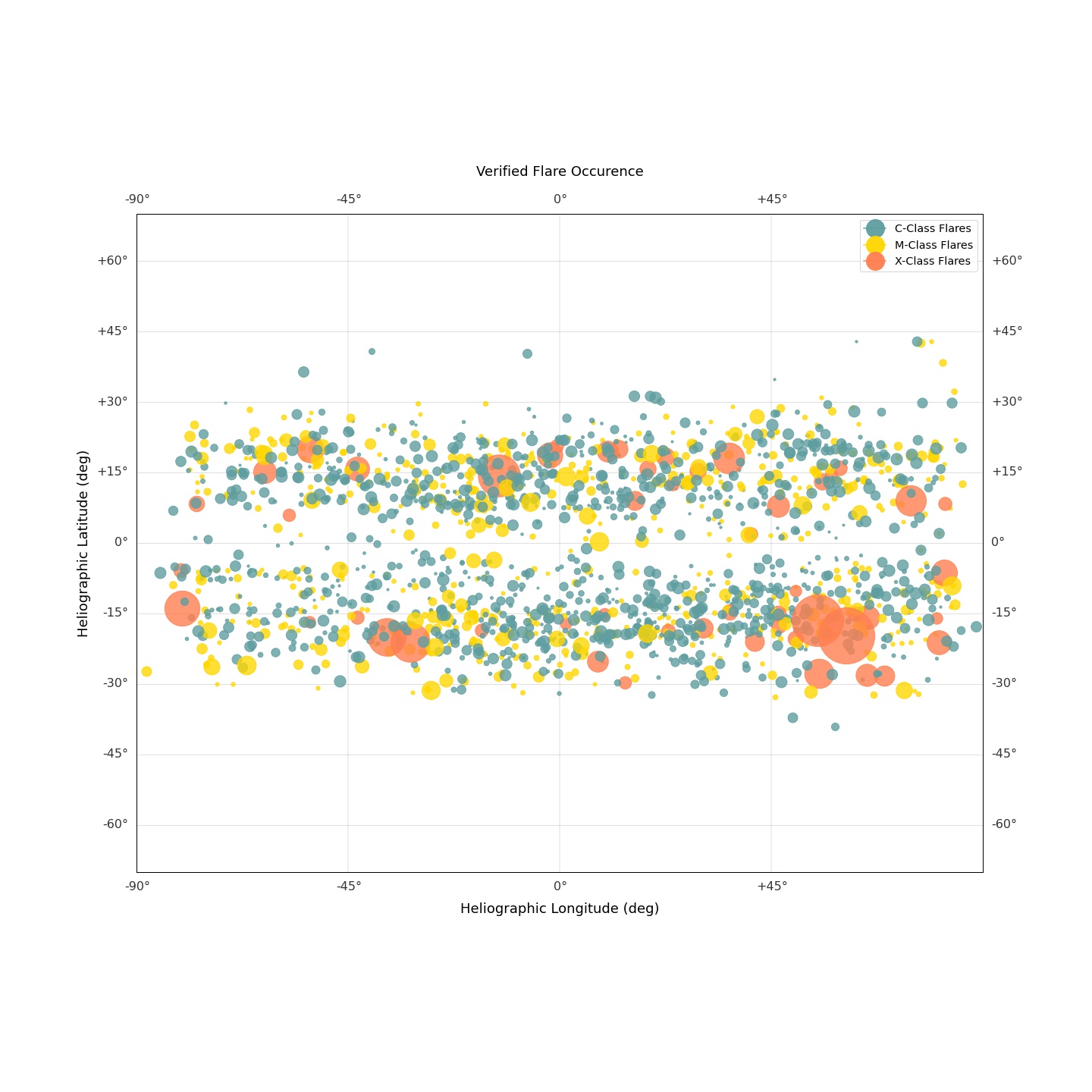}
    \caption{Distribution of all the verified EIT flares on heliographic coordinates.}
    \label{fig:5}
\end{figure}

\begin{figure}[ht!]
    \centering
    \includegraphics[width=0.9\textwidth,trim={0 10cm 0 10.5cm},clip]{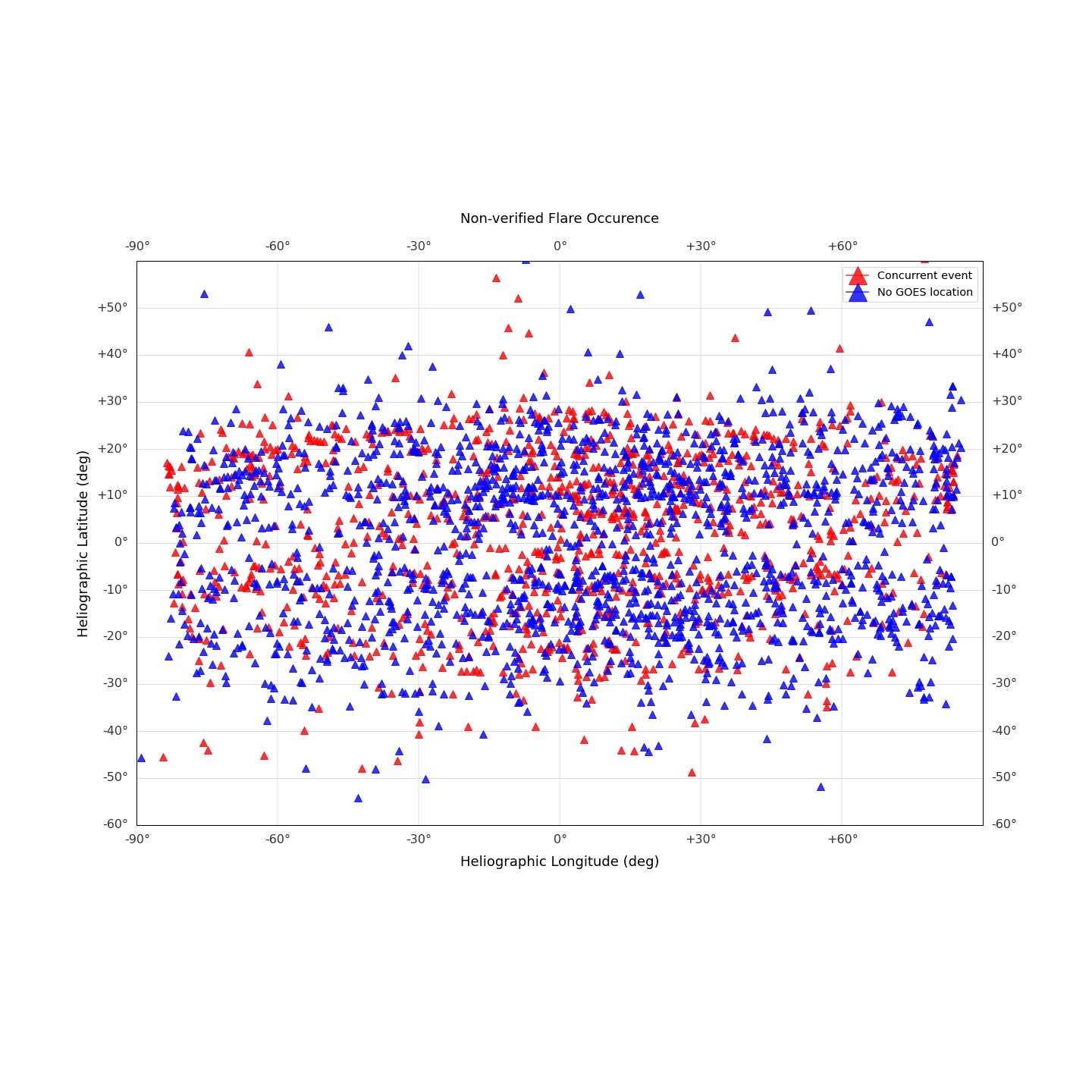}
    \caption{Distribution of all the non-verified EIT flares on heliographic coordinates.}
    \label{fig:6}
\end{figure}

\subsubsection{Statistics}\label{subsec:stat}

Figure \ref{fig:4} shows a bar plot of the EIT flares from Jan 1997 to Aug 2010. Overall, we have obtained 4602 flares from the EIT data. In terms of numbers, 1881 flares are matching spatially and temporally with GOES flares, which is considered verified, while 1083 EIT flares match temporally. However, 1638 flares do not have a corresponding GOES location. Figure \ref{fig:9} shows a comparison between the number of flares in the GOES catalog for GOES class $>$C1.0 with and without coordinate values, respectively. The number of flares with no locations in the GOES catalog dominates for all the years except 2003. This gap greatly influences the verification process as a result, and the ratio of verified to non-verified EIT flares is sharply reduced. However, the overall ratio is similar between the verified EIT flares and the GOES flares with a location. It remains between $\approx$ 20\% to $\approx$ 40\% from 1997 to 2003, while it is $>$ 40\% from 2004 to 2006 and 2010. The ratio falls below 20\% for 2008 and 2009. Figure \ref{fig:7} compares the total number of GOES and EIT flares for our analysis. Figures \ref{fig:5} and \ref{fig:6} show scatter plots of all the verified and non-verified EIT flares, respectively, based on the GOES classes. Figure \ref{fig:5} shows scatter plot of all the verified flares in accordance to the GOES classes. The two sub-types of \textit{non verified} flares are shown in Figure \ref{fig:6} but are not represented based on flare intensities as they cannot be confirmed. Also seen are some of the flares occurring near-limb.

\begin{deluxetable*}{ccccc}
\tablenum{2}
\tablecaption{The number of flares in the GOES catalog verified between 1997 and 2017 using secondary sources. \label{tab2}}
\tablewidth{0pt}
\tablehead{
\colhead{\textbf{Source}} & \colhead{\textbf{Period}} & \colhead{\textbf{C-Class}} & \colhead{\textbf{M-Class}} & \colhead{\textbf{X-Class}}
}
\startdata
\hline
SOHO/EIT & 1997-01-01 to 2010-06-30 & 1182 & 647 & 52\\
SDO/AIA \citep{dml}  & 2010-05-01 to 2018-09-01 & 6319 & 694 & 50      
\enddata
\end{deluxetable*}

Statistical analysis shows that 41$\%$ of the EIT flares are verified with the GOES catalog while 35.5$\%$ of the EIT flares do not have a corresponding GOES location. The remaining 23.5$\%$ of the EIT flares do not match spatially with GOES. Our statistics show satisfactory results despite many challenges, as stated earlier. We present a summary of our verified EIT flare findings in Table \ref{tab2} along with the SDO data from \citet{dml}.

\section{Conclusions} \label{sec:conc}

The SOHO/EIT flare catalog for 195 {\AA} is developed for  predictive and exploratory analytics using ML in flare prediction and to the scientific community for investigation of flares during the Solar Cycle 23. This catalog primarily serves to verify the locations of flares reported in the GOES catalog as a first step to generate a benchmark dataset used for flare prediction. Here, verification implies the spatiotemporal matching of the flaring events between the GOES and EIT data. However, some events are not verified because (i) the GOES catalog has no report of exclusive coordinate locations for many flares in order to cross-check and (ii) multiple events are detected in the EIT images. Altogether, we identified 4602 events from the EIT images between January 1997 and July 2010, out of which 1881 flares overlap with the GOES catalog in temporal and spatial coordinates. The SOHO/EIT Flare Catalog will be publicly available on Harvard dataverse for the solar physics community in \url{https://doi.org/10.7910/DVN/C9H34R}. We appreciate any suggestions concerning the catalog.

As far as the combining of EIT and GOES data is concerned, our foremost goal is to link the GOES flares with the correct magnetogram locations. So, the preflare magnetogram data can be used for testing flare prediction algorithms. Furthermore, it has become clear that there may be signs in preflare EUV emission predicting an upcoming flare.  We can now verify those signatures from a large database of the EIT and Atmospheric Imaging Assembly (AIA) observed flares.

\acknowledgments
S.R. thanks the members of DM Lab, Dr. Georgoulis and Ms. Aparna Venkataramanasastry, for the many discussion during the project. This work was supported in part by funding from NASA Space Radiation Analysis Group (under GSU's Sponsor Award No. TXS0156755). All scientific images used in this work are courtesy of the SOHO/EIT Consortium; SOHO is a joint ESA-NASA program. We are also very grateful to the anonymous reviewer who provided constructive feedback that greatly improved the quality and scope of this paper. 






\bibliography{sample63}{}
\bibliographystyle{aasjournal}

%%\allauthors

\listofchanges
\end{document}